\documentclass[twocolumn,english,showpacs,superscriptaddress,prc]{revtex4-1}
\usepackage[T1]{fontenc}
\usepackage[latin9]{inputenc}
\usepackage{babel}
\usepackage{booktabs}
\usepackage{amsmath}
\usepackage{amsthm}
\usepackage{amssymb}
\usepackage{graphicx}
\usepackage{wasysym}
\usepackage[unicode=true,pdfusetitle,
 bookmarks=true,bookmarksnumbered=false,bookmarksopen=false,
 breaklinks=false,pdfborder={0 0 1},backref=false,colorlinks=false]
 {hyperref}
\usepackage{breakurl}
\usepackage{color}



\begin{document}

\title{Quark number scaling of $p_{T}$ spectra for $\Omega$ and $\phi$ in relativistic heavy-ion collisions}

\author{Jun Song }
\affiliation{Department of Physics, Jining University, Shandong 273155, China}

\author{Feng-lan Shao}
\email{shaofl@mail.sdu.edu.cn}
\affiliation{School of Physics and Physical Engineering, Qufu Normal University, Shandong 273165, China}

\author{Zuo-tang Liang}
\email{liang@sdu.edu.cn}
\affiliation{Key Laboratory of Particle Physics and Particle Irradiation (MOE), Institute of Frontier and Interdisciplinary Science, Shandong University, Qingdao, Shandong 266237, China}

\begin{abstract}
    We show that the experimental data of transverse momentum ($p_{T}$) spectra of $\Omega$ baryon and $\phi$ meson at mid-rapidity in heavy-ion collisions exhibit the constituent quark number scaling in a wide energy range from RHIC to LHC.  Such a scaling behavior is a direct consequence of quark combination mechanism via equal velocity combination and provides a very convenient way to extract the $p_{T}$ spectrum of strange quarks at hadronization.  We present the results of strange quarks obtained from the available data and study the properties in particular the energy dependence of the averaged transverse momentum $\langle p_{T}\rangle$ and the transverse radial flow velocity $\langle\beta\rangle$ with a hydrodynamics-motivated blast-wave model.
\end{abstract}

\maketitle


\section{Introduction}
\label{introduction}
Precision measurements on transverse momentum ($p_T$) spectra of hadrons 
produced in high energy $pp$, $pA$ and $AA$ collisions 
at the Relativistic Heavy Ion Collider (RHIC) and the Large Hadron Collider (LHC) 
in the intermediate $p_T$ region provide us a good opportunity to study 
the hadronization mechanism and in particular an efficient probe to 
the created quark matter system --- the quark gluon plasma (QGP).   
We have in particular data for hadrons such as $\Omega^{-}(sss)$ hyperon 
and $\phi(s\bar{s})$ \cite{ Adams:2006ke,Abelev:2008aa,ABELEV:2013zaa,Abelev:2014uua,Zhu:2014xga, Adamczyk:2015lvo,Aggarwal:2010ig,Agakishiev:2011ar,Abelev:2008zk} that consist of only strange quarks and/or anti-quarks and 
are important probe to strangeness related dynamics of QGP in $AA$ 
collisions~\cite{Shor:1984ui,vanHecke:1998yu,Singh:1986ia,Bass:1999tu,Dumitru:1999sf,Chen:2008vr}.
Because they are expected to have relatively small hadronic interaction
cross sections~\cite{Shor:1984ui,vanHecke:1998yu}, 
they suffer from small distortion in hadronic re-scattering stage and therefore carry important 
information of QGP at hadronization.

In a recent work~\cite{Song:2017gcz}, we showed that the experimental data of $p_{T}$ spectra
of hadrons at mid-rapidity in high-multiplicity events of  $p$-Pb collisions at LHC energies exhibit a perfect
constituent quark number scaling (QNS). 
Such a scaling behavior is a direct consequence of quark combination mechanism via equal velocity combination (EVC) 
and might be regarded as a clear signature for creation of QGP. 
Recently, QNS for hadronic $p_T$ spectra has shown also valid in 
high-multiplicity $pp$ collisions at $\sqrt{s}=$7 and 13 TeV~\cite{Gou:2017foe,Zhang:2018vyr}.
It is nature to ask whether it is also valid in $AA$ collisions. 

The study in $AA$ collisions is in principle straightforward.  
However since in the intermediate $p_T$ region, for long lived hadrons such as pions, Kaons and protons,  
decay contributions are often important and contaminations from these decays 
and final-state hadronic interactions are difficult to remove. 
It is therefore very exciting to see that data on  $\Omega$ and $\phi$  have been 
obtained~\cite{ABELEV:2013zaa,Abelev:2014uua,
Adams:2006ke,Zhu:2014xga,Abelev:2008aa,Aggarwal:2010ig,Agakishiev:2011ar,Abelev:2008zk,Adamczyk:2015lvo} 
at RHIC and LHC in a very wide energy range from $\sqrt{s_{NN}}=11.5$ to $2760$GeV. 

The $p_{T}$ spectra of $\Omega$ and $\phi$ provide the best place to test QNS not only because  
there is little contamination from decay but also due to the fact that in these hadrons only 
constituent strange quarks and anti-quarks are involved so that QNS, if exists, takes the simplest form. 
It provides also an ideal place to extract the $p_T$ spectrum for strange quarks. 

In this paper, we examine the experimental data of mid-rapidity
$p_{T}$ spectra of $\Omega$ and $\phi$ in heavy-ion collisions~\cite{ABELEV:2013zaa,Abelev:2014uua,
Adams:2006ke,Zhu:2014xga,Abelev:2008aa,Aggarwal:2010ig,Agakishiev:2011ar,Abelev:2008zk,Adamczyk:2015lvo}
and show that such a QNS also exists in the broad energy region from RHIC to LHC. 
We extract the strange quark $p_T$ spectrum just before
hadronization in relativistic heavy-ion collisions and study the related properties 
within a hydrodynamics-motivated blast-wave model. 
These results are given in Secs. II and III. 
In Sec. IV, we present a short summary and an outlook. 

\section{The QNS for hadronic $p_{T}$ spectra}

QNS was shown to be valid for hadronic $p_T$ spectra in high-multiplicity events
in $pp$ and $p$-Pb collisions at LHC~\cite{Song:2017gcz, Zhang:2018vyr}.
We now examine whether it is also valid in $AA$ collisions by using data on $p_T$ spectra
obtained at RHIC and LHC~\cite{ABELEV:2013zaa,Abelev:2014uua,
Adams:2006ke,Zhu:2014xga,Abelev:2008aa,Aggarwal:2010ig,Agakishiev:2011ar,Abelev:2008zk,Adamczyk:2015lvo}.

\subsection{QNS for $\Omega^{-}$ and $\phi$ in $AA$ collisions}

\begin{figure*}[htb]
\begin{centering}
\includegraphics[scale=0.8]{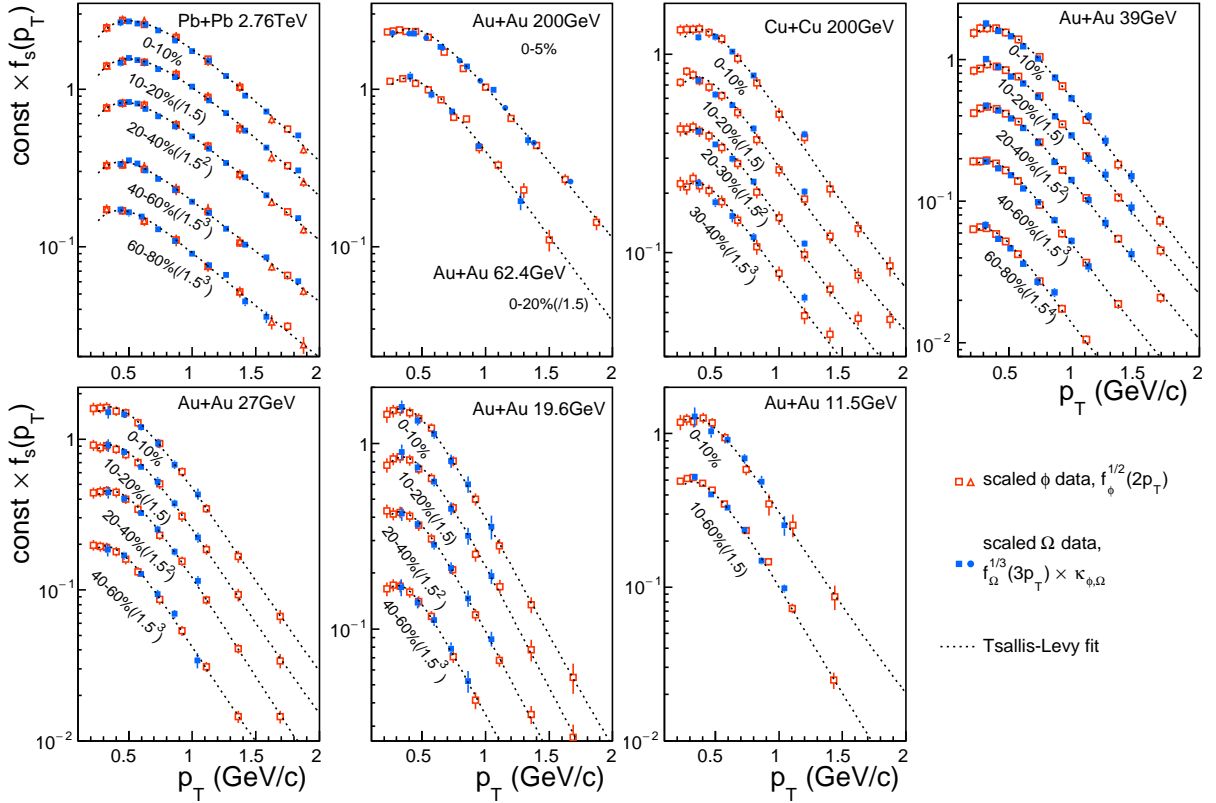}
    \caption{Test of QNS of $p_T$ spectra of $\Omega$ ($\Omega^{-}+\bar{\Omega}^{+}$) and $\phi$ in $AA$ collisions.
The constant $\kappa_{\phi,\Omega}$ is adjusted in the way
that the results for $\kappa_{\phi,\Omega}f_{\Omega}^{1/3}(3p_{T})$
look to fall in the same lines as those of $f_{\phi}^{1/2}(2p_{T})$.
The data are taken from~\cite{ABELEV:2013zaa,Abelev:2014uua,Adams:2006ke,
Zhu:2014xga,Abelev:2008aa,Aggarwal:2010ig,Agakishiev:2011ar,Abelev:2008zk,Adamczyk:2015lvo}.
The dotted lines are simple fittings with the Tsallis-Levy function~\cite{Tsallis1988} to guide the eye. \label{fig1}}
\par\end{centering}
\end{figure*}

We recall that QNS for hadronic $p_T$ spectra in $pp$ and $p$-Pb collisions at LHC
in the intermediate $p_T$ region refers to the number of constituent quarks and
is formulated in the following way.
For $p_T$ spectra $f_h(p_{T})\equiv dN_h/dp_T$ of hadrons consisting of quarks and/or anti-quarks of the same flavor,
we have,
\begin{equation}
f_h(p_{T})=\kappa_h f_q^{n_q}(p_{T}/n_q),\label{eq:fh_qns}
\end{equation}
where $h$ denotes hadron, $n_q$ is the number of the constituent quarks and/or anti-quarks,
$\kappa_h$ is a constant independent of $p_T$ but can be different for different hadron $h$.
For a hadron consisting of different flavors of quarks, e.g., for a baryon $B$ consisting of $q_1q_2q_3$,
we have,
\begin{equation}
f_B(p_{T})=\kappa_B f_{q_1}(x_1p_{T})f_{q_2}(x_2p_{T})f_{q_3}(x_3p_T), \label{eq:fb}
\end{equation}
 where $x_1+x_2+x_3=1$, $x_1:x_2:x_3=m_{q_1}:m_{q_2}:m_{q_3}$ and
 $m_{q_i}$ is the constituent quark mass of $q_i$.
 The function $f_{q_i}(p_T)$ of $p_T$ is universal for different hadrons
 and can be identified as the $p_T$-distribution of $q_i$ before hadronization.

 It is obvious that for $\Omega^{-}$ and $\phi$, QNS takes the simplest form, i.e.,
 \begin{align}
&f_{\Omega^{-}}(3p_{T})  =\kappa_{\Omega^{-}}f_{s}^3(p_{T}),\label{eq:fpt_Omega}\\
&f_{\phi}(2p_{T})  =\kappa_{\phi} f_{s}(p_{T})f_{\bar{s}}(p_{T})=\kappa_{\phi}f_{s}^2(p_{T}), \label{eq:fpt_phi}
\end{align}
and it leads to the equality,
\begin{equation}
  f_{\phi}^{1/2}(2p_{T})=\kappa_{\phi,\Omega^{-}}f_{\Omega^{-}}^{1/3}(3p_{T}), \label{eq:qns}
\end{equation}
where the coefficient $\kappa_{\phi,\Omega^{-}}=\kappa_{\phi}^{1/2}/\kappa_{\Omega^{-}}^{1/3}$
is a constant that is independent of $p_{T}$ but can be dependent on collision energies, centralities or other parameters.
Here we consider the mid-rapidity region. 
Since all strange quarks and anti-quarks are produced in pairs in the collision process, 
we take as usual $f_{s}(p_T)=f_{\bar{s}}(p_T)$ that was used to provide good description of 
strange and anti-strange hadron $p_T$ spectra at RHIC energies~\cite{Shao:2009uk,Sun:2011kj, Wang:2012cw}.

Eq.~(\ref{eq:qns}) can be used to check whether QNS is valid for the $p_T$ spectra of  $\Omega^{-}$ and $\phi$ in AA collisions.
To do this, we take all of the available data on the $p_T$ spectra of $\Omega$ ($\Omega^{-}+\bar{\Omega}^{+}$)
and $\phi$ in Au+Au collisions at RHIC and in Pb+Pb collisions at LHC~\cite{ABELEV:2013zaa,Abelev:2014uua,
Adams:2006ke,Zhu:2014xga,Abelev:2008aa,Aggarwal:2010ig,Agakishiev:2011ar,Abelev:2008zk,Adamczyk:2015lvo}.
We build up $f_{\phi}^{1/2}(2 p_{T})$ and $f_{\Omega}^{1/3}(3 p_{T})$ and plot them in the same figure.
The obtained results are given in Fig.~\ref{fig1}.
The results for $\Omega$'s are multiplied by an arbitrary $p_{T}$-independent constant
that corresponds to $\kappa_{\phi,\Omega}$ in Eq.~(\ref{eq:qns}).
This constant is adjusted in the way
that the results for $\kappa_{\phi,\Omega}f_{\Omega}^{1/3}(3p_{T})$ look to fall in the same lines as those of $f_{\phi}^{1/2}(2p_{T})$.
To guide the eye, we also present simple fittings with the Tsallis-Levy function~\cite{Tsallis1988} in the figure.

From Fig.~\ref{fig1}, we see clearly that  the results obtained from the data of $\Omega$ and those of $\phi$
are well coincident with each other at different energies and different centralities.
These results clearly show that QNS exists also in heavy-ion collisions in such a wide energy range.
Besides the normalization constant, the curve in Fig.~\ref{fig1} just corresponds to $f_s(p_T)$,
the $p_T$ spectrum of strange quarks before hadronization.

\subsection{QNS and quark combination via EVC}
\label{secQNSEVC}

It has been shown that QNS for hadronic $p_T$ spectra is a direct consequence of
the quark combination via equal velocity combination (EVC).
The reason is in fact very simple.
If we demand that, e.g., a baryon $B$ consisting of $q_1q_2q_3$ is produced
through combination of the three constituent quarks with the same velocity $\beta$,
we obtain immediately from $p_i=\gamma\beta m_i$ that $p_1:p_2:p_3=m_1:m_2:m_3$
and the results given by Eqs.~(\ref{eq:fb})-(\ref{eq:fpt_phi}).
The basic idea of EVC in the quark combination mechanism can be traced back to~\cite{Lin:2003jy} 
and recently inspired by QNS has been applied to describe 
hadronic $p_T$ spectra in different situations~\cite{Gou:2017foe,Song:2018tpv,Li:2017zuj,Zhang:2018vyr}.
It is also in this mechanism, the universal $f_{q_i}(p_T)$ gets a clear physical significance.
It is just the $p_T$ spectrum of $q_i$ before hadronization.
In other words, we obtain a simple way to extract the $p_T$ spectra of quarks before hadronization.

The value of the coefficient $\kappa_h$ depends on
the dynamical details in the combination such as the baryon to meson yield ratio,
vector to pseudo-scalar meson production ratio,
strangeness suppression, net quark influence and so on. 
It can not derived in QNS but should be derived from the hadronization model.
In the limiting case that there is no net quark contribution and
the numbers of quarks and anti-quarks are very large,
it has been shown~\cite{Song:2017gcz} that, in quark combination models,
$\kappa_h$ takes a relatively simple form, i.e.,
\begin{align}
\label{eq:kappaOmega}
    \kappa_{\Omega}^{(0)} &= \frac{8}{3 a} A_{sss}/ N_t^{2}, \\
 \label{eq:kappaphi}
    \kappa_{\phi}^{(0)} &= 2 (1-\frac{1}{a}) A_{ss} C_{\phi}/ N_t,
\end{align}
where a superscript $(0)$ is introduced to specify that it is for zero net quark,
$N_t=N_{q}+N_{\bar{q}}$ is the total number of quarks and antiquarks;
$a$ is a parameter describing the baryon to meson production ratio
that was parameterized as~\cite{Song:2013isa} $N_{M}/N_{B}=3(a-1)$ in quark combination models;
$C_{\phi}$ is the conditional probability to produce a $\phi$ meson from a $s\bar{s}$
under the condition that the $s\bar{s}$ is definite to combine with each other to form a meson.
$A_{ss}^{-1}\equiv 2\int dp_{T}f_{s}^2(p_{T})$
and $A_{sss}^{-1}\equiv 3\int dp_{T}f_{s}^3(p_{T})$
are determined by integrations of the normalized $s$-quark $p_T$-distribution $f_{s}(p_{T})$ squared and cuboid respectively.

From Eqs.~(\ref{eq:kappaOmega}) and (\ref{eq:kappaphi}), we obtain immediately,
\begin{equation}
    \kappa_{\phi,\Omega}^{(0)}  = \frac{3^{1/3}}{\sqrt{2}}\frac{\sqrt{a-1}}{a^{1/6}}\,\frac{A_{s{s}}^{1/2}}{A_{sss}^{1/3}}\,C_{\phi}^{1/2} N_t^{1/6}.
 \label{eq:kappa_phi_Omega_z0}
\end{equation}
We generalize it to the case of finite quark numbers and non-zero net-quark numbers~\cite{Song:2017gcz,Gou:2017foe}
and obtain the following result,
up to order of $N_t^{-1}$ and $z^{2}$,
\begin{align}
    \kappa_{\phi,\Omega}  \approx
    \kappa_{\phi,\Omega}^{(0)} &{} \left[1+\frac{2\left(2+\lambda_{s}\right)}{(1-z)\lambda_{s}N_t}     -\frac{1}{18}(a-4)(2a-7)z^{2} \right],
    \label{eq:kappa_phi_Omega}
\end{align}
where $\lambda_{s}\equiv N_{\bar s}/N_{\bar{u}}$ is the strangeness suppression factor and
$z\equiv\left(N_{q}-N_{\bar{q}}\right)/N_t$ is the net-quark fraction in the system.
We emphasize in particular that net quark can only be $u$ or $d$ quark so that it has no 
influence on the strangeness conservation.

\begin{figure}[htb]
\centering{}\includegraphics[scale=0.42]{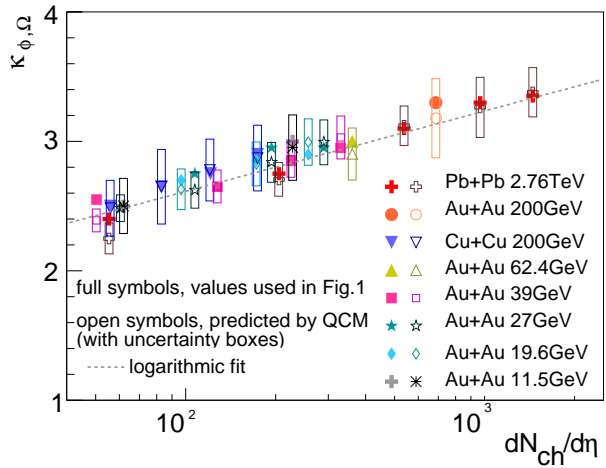}
    \caption{Coefficient $\kappa_{\phi,\Omega}$ used in Fig. \ref{fig1}, the full symbols,
and the predicted values in quark combination model, open symbols
with uncertainty boxes. \label{fig2}}
\end{figure}

The result given by Eq.~(\ref{eq:kappa_phi_Omega}) is valid generally in the stochastic quark combination models.
It depends on the properties of quark composition of the system before hadronization such as the total quark number, 
the net quark fraction and the strangeness suppression, 
and also on the hadronization mechanism via baryon to meson ratio and the conditional probability $C_{\phi}$. 
These parameters empirically known and are extracted from the corresponding data or other empirical facts.
The total number of quarks and anti-quarks $N_t$ and the net-quark fraction $z$ are determined
by the charged hadron multiplicity and anti-hadron to hadron yield ratios (see e.g., \cite{Shao:2009uk,Song:2013isa,Shao:2015rra}).
The parameter $a$ is determined as $a=4.86$ in the light-flavor sector~\cite{Shao:2017eok}.
The strange suppression factor $\lambda_{s}$ is determined by yields of strange
hadrons kaons and $\Lambda$'s relative to pions~\cite{Shao:2009uk,Shao:2017eok}.
If we take only vector and pseudo-scalar meson productions into account,
$C_{\phi}$ is determined by vector to pseudo-scalar meson production ratio.
It can also be determined by the data
of the yield ratio $\phi/K^{-}$~\cite{Abelev:2008aa,Abelev:2008zk,Abelev:2014uua,Adamczyk:2015lvo}
using the relation $\phi/K^{-}=\lambda_{s}C_{\phi}/\left(1+\lambda_{s}C_{\phi}\right)$ in the quark combination model.
As a test, we show in Fig.~\ref{fig2} values of coefficient $\kappa_{\phi,\Omega}$
(full symbols) obtained in Fig.~\ref{fig1} and those obtained using Eq.~(\ref{eq:kappa_phi_Omega}) (open symbols).
The model uncertainties come from those of the parameters, in particular those of $C_{\phi}$ and $N_t$.
We see that the agreement with each other is quite satisfactory 
that suggests that not only the shapes of $p_T$-spectra but also yields of $\Omega$ and $\phi$ can be described 
by the combination model under EVC. 
We also see that there is a very good agreement with a logarithmic fit for the coefficient $\kappa_{\phi,\Omega}$
as a function of $dN_{ch}/d\eta$.

\section{The $p_T$ spectrum of $s$-quarks}
\label{secPT}

As described in Sec.~II, QNS provides a convenient way of extracting
$p_T$ spectra for quarks before hadronization.
Here, from the data on $\Omega$ and $\phi$ \cite{ABELEV:2013zaa,Abelev:2014uua,
Adams:2006ke,Zhu:2014xga,Abelev:2008aa,Aggarwal:2010ig,Agakishiev:2011ar,Abelev:2008zk,Adamczyk:2015lvo},
we obtain the $p_T$ spectrum of strange quarks.
It is then interesting to study the related properties based on the extracted results.
In this section, we present the results on the energy dependence of the averaged transverse momentum
and the radial flow velocity within the blast-wave model~\cite{Schnedermann:1993ws}.

\subsection{The energy dependence of $\langle p_T\rangle$}
\label{secPTave}

\begin{figure}[htb]
\centering{}\includegraphics[scale=0.4]{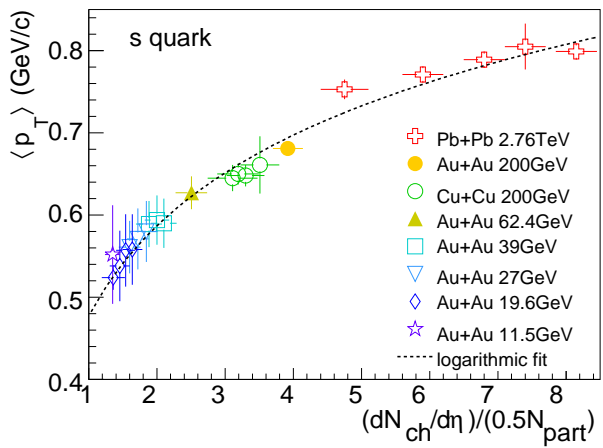}
\caption{The average transverse momentum $\left\langle p_{T}\right\rangle $ of strange quarks
at mid-rapidity in heavy-ion collisions.
The dashed line is a logarithmic fit. \label{fig3}}
\end{figure}

By fitting the data of $\Omega$ and $\phi$ as given in Fig.~\ref{fig1}
in terms of the Tsallis-Levy function~\cite{Tsallis1988} of quark $p_T$ distribution,
we calculate the averaged transverse momentum $\left\langle p_{T}\right\rangle$
of strange quarks in the soft region $0<p_{T}<2$ GeV/c.
The results obtained are shown in Fig.~\ref{fig3}
as a function of the charged-particle pseudo-rapidity density
per pair of participant nucleons $\left(dN_{ch}/d\eta\right)/\left(0.5N_{part}\right)$.
The error bars are calculated from the uncertainties of the parameters
of Tsallis-Levy function in fitting the data with quadratic sums of statistical and systematic uncertainties.
The dashed line represents a logarithmic fit
$\langle p_T\rangle=0.476+0.159 \ln[(dN_{ch}/d\eta)/(0.5N_{part})]$ GeV.

From Fig.~\ref{fig3}, we see that the transverse momentum $\langle p_T\rangle$
for $s$-quarks before hadronization depends approximately logarithmically on $\left(dN_{ch}/d\eta\right)/\left(0.5N_{part}\right)$.
The energy dependence is not significant.
We note that in particular those results at $\sqrt{s_{NN}}=$19.6 and 11.5 GeV
are mixed up each other due to the similar $\left(dN_{ch}/d\eta\right)/\left(0.5N_{part}\right)$.
The averaged transverse momentum of strange quarks
under local thermal equilibrium~\cite{Schnedermann:1993ws} is about 0.45-0.5 GeV/c
at the hadronization temperature ($T\approx$160 MeV).
The obtained $\left\langle p_{T}\right\rangle$ of strange quarks in Fig.~\ref{fig3} is significantly larger than this value.
This might suggest a large collective transverse radial flow of strange quarks created in prior parton phase evolution.

\subsection{The radial flow velocity of strange quarks within the blast-wave model}
 \label{secRadial}

Using the $p_T$-spectrum obtained in Sec.~\ref{secPTave},
we can further analyze the radial flow of strange quarks
within the hydrodynamics motivated blast-wave model~\cite{Schnedermann:1993ws}.
Here, in this model, it is envisaged that particles are locally thermalized and moving with a collective
transverse radial flow velocity field.
The $p_{T}$-distribution is obtained from the superposition of
boosted thermal sources at temperature $T$, i.e.~\cite{Schnedermann:1993ws},
\begin{equation}
\frac{dN}{p_{T}dp_{T}}\propto\int_{0}^{1}\xi d\xi\,m_{T}I_{0}\left(\frac{p_{T}\sinh\rho}{T}\right)K_{1}\left(\frac{m_{T}\cosh\rho}{T}\right), \label{eq:bw}
\end{equation}
where $\rho=\tanh^{-1}\beta\left(\xi\right)$ is the boost angle,
$I_{0}$ and $K_{1}$ are modified Bessel functions, and $m_{T}=\sqrt{m^{2}+p_{T}^{2}}$ is the transverse mass.
The flow velocity profile is taken as $\beta\left(\xi\right)=\beta_{s}\xi^{n}$
where $\xi$ is the relative radial position, the form of the profile is controlled by the exponent $n$,
and $\beta_s$ is the surface velocity.
To be compatible with hydrodynamics simulations, we set $n$ to $n\leq2$.
We take the temperature $T$, the averaged flow velocity $\langle\beta\rangle=2\beta_{s}/(n+2)$
and the exponent $n$ as fit parameters.

\begin{figure}[htbp]
\centering{}\includegraphics[scale=0.4]{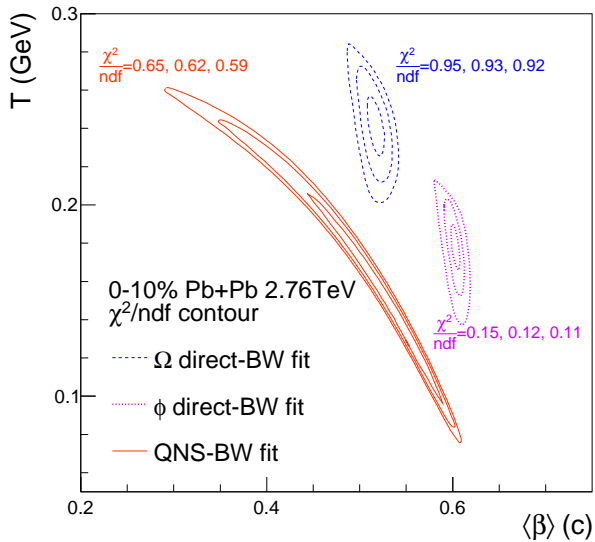}
\caption{Contour plot of blast-wave model fit of data of the $p_T$ spectra of  $\Omega$ and $\phi$ in central (0-10\%)
Pb+Pb collisions at $\sqrt{s_{NN}}=$ 2.76 TeV~\cite{ABELEV:2013zaa,Abelev:2014uua}.
\label{fig4}}
\end{figure}

We take the data on $p_T$-spectra of $\Omega$ and $\phi$ in central (0-10\%) Pb+Pb collisions
at $\sqrt{s_{NN}}=$ 2.76 TeV~\cite{ABELEV:2013zaa,Abelev:2014uua} as an example,
and show the fit results of blast-wave model in ($\langle\beta\rangle$-$T$) plane in Fig.~\ref{fig4}.
For comparison, we first make the direct fit at the hadron level for
$\Omega$ ($p_{T}\leq6$ GeV/c) and $\phi$ ($p_{T}\leq4$ GeV/c) respectively,
then make the fit for the strange quark distribution $f_{s}(p_{T})$ obtained in Fig.~\ref{fig1}.
The results obtained are marked as direct-BW or QNS-BW fit in the figure.

From Fig.~\ref{fig4}, we see that the center region of the direct-BW fit to $\Omega$ is far away from that to $\phi$.
Compared with those for $\phi$-mesons, the freeze-out temperature of $\Omega$
obtained this way is higher and the velocity is smaller.
This would imply that $\Omega$ might freeze out earlier than $\phi$ in central Pb+Pb collisions at $\sqrt{s_{NN}}=$ 2.76 TeV.
However, if we do similar fits for data in central Au+Au collisions at $\sqrt{s_{NN}}=200$ or $39$ GeV~\cite{Adams:2006ke,
Zhu:2014xga,Abelev:2008aa,Adamczyk:2015lvo},
we arrive at different conclusions in different cases:
At $\sqrt{s_{NN}}=200$ GeV,
the center region of the fit to $\Omega$ is lower in temperature and higher in $\langle\beta\rangle$
(which is consistent with the previous report of STAR collaboration~\cite{Adams:2005dq}),
while at 39 GeV, it is lower in temperature and lower in $\langle\beta\rangle$.
In other collision energies, the relative position between $\Omega$ and $\phi$ is also variant.
This indicates that one could have difficulty to obtain a consistent set of freeze-out parameters for $\Omega$ and $\phi$ 
by individual blast-wave model fits. 
In this case, one can only rely on the combined fit for both of them to obtain a common parameter space describing 
the available data within the accuracy achieved yet.  

On the other hand, as inspired by QNS, it could be also interesting to attribute the production of $\Omega$ and $\phi$ 
to a common source of strange quarks and anti-quarks at hadronization, and apply the blast-wave model to quark level.
From Fig.~\ref{fig4}, we see that a characteristic feature of the parameter space of strange quarks
is a narrow band across relatively large $\langle\beta\rangle$ or $T$ range in the plane.
This is because, in contrast to $\Omega$ and $\phi$, strange quark has a small mass
that has small influence on the shape of the distribution given by Eq.~(\ref{eq:bw}).
The $p_T$-spectrum of strange quarks is not very sensitive to $T$ itself but 
determined by $\langle\beta\rangle$ and $T$ in a complementary manner. 
To determine the $\langle\beta\rangle$ or $T$, one needs to invoke inputs from studies on other aspects of the quark system.
Compared to the direct fits to $\Omega$ and/or $\phi$, the parameter space of strange quarks
has a shift toward to smaller $\langle\beta\rangle$ direction.
We observe the same behavior by fitting the data~\cite{Adams:2006ke,
Zhu:2014xga,Abelev:2008aa,Aggarwal:2010ig,Abelev:2008zk,Adamczyk:2015lvo} at other collision energies.

Such a shift is the consequence of quark combination and can be demonstrated by a simple example. Suppose  quark distribution in the rest frame has a Boltzmann form  $dN_{s}/( p_Tdp_T)=f_s(p_T)/p_T\propto\exp(-\sqrt{p_T^2+m_s^2}/T)$ in two-dimensional transverse space, we obtain $f_{\phi}(p_T)/p_T\propto p_T\exp(-\sqrt{p_T^2+m_{\phi}^2}/T) $ and $f_{\Omega}(p_T)/p_T\propto p_T^2\exp(-\sqrt{p_T^2+m_{\Omega}^2}/T)$ by applying Eqs. (\ref{eq:fpt_Omega}) and (\ref{eq:fpt_phi}) and using $m_{\phi}\approx 2m_s$ and $m_{\Omega}\approx 3 m_s$. We see that $f_{\phi}(p_T)$ and  $f_{\Omega}(p_T)$ in the rest frame are broader than Boltzmann distribution by an extra $p_T$ and $p_T^2$, respectively. This extra $p_T$ dependence will cause larger temperature and/or flow velocity in above direct BW-fit for $\Omega$ and $\phi$.

\begin{figure}[htbp]
\centering{}\includegraphics[scale=0.40]{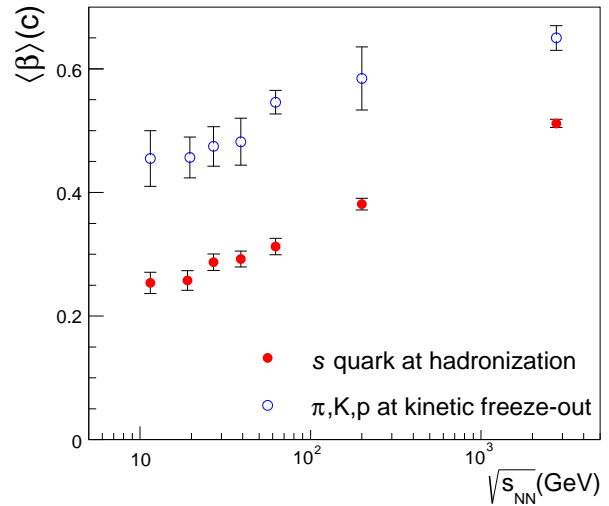}
\caption{The averaged radial flow velocity $\langle\beta\rangle$ of strange quarks at hadronization in central heavy-ion collisions extracted 
from $p_{T}$ spectrum data of $\Omega$ and $\phi$ at mid-rapidity,  compared with those obtained for fitting the $\pi$, $K$ and protons data~\cite{Abelev:2013vea,Adamczyk:2017iwn}. \label{fig5}}
\end{figure}

To see the qualitative behavior of $\langle\beta\rangle$ for strange quarks at hadronization, we simply take $T=T_{0}(1-c_{2}\mu_{B}^{2}/T_{0}^{2})$,
where $\mu_{B}$ is baryon number chemical potential and is taken as~\cite{Andronic:2017pug}
$\mu_{B}=1.3075/(1+0.288\sqrt{s_{NN}})$ GeV;
 the curvature $c_{2}=0.0145$ is taken from Lattice QCD calculations~\cite{Bonati:2018nut};
and $T_{0}$ is the temperature at vanishing $\mu_{B}$
and is taken as $T_{0}=164\pm 5$ MeV~\cite{Cleymans:2005xv,Bellwied:2013cta,Chatterjee:2013yga,ABELEV:2013zaa,Shao:2015rra,Andronic:2017pug,Becattini:2017pxe}.
Fig.~\ref{fig5} shows the results obtained in central collisions at different collision energies.
For comparison, we show also results for direct fit to
the data of $\pi$, $K$ and $p$'s~\cite{Abelev:2013vea,Adamczyk:2017iwn},
which characterize the averaged radial flow at kinetic freeze-out.
We see that $\langle\beta\rangle$ of strange quarks increases monotonically with increasing energy.
Compared with those for $\pi$, $K$ and protons, the difference seems to be smaller at the LHC energy
implying smaller contributions from hadronic stage.

\section{Summary and outlook}

To summarize, we show that the experimental data of $p_{T}$ spectra of $\Omega$
and $\phi$ in $AA$ collisions at both RHIC and LHC energies exhibit also the quark number scaling (QNS).
This suggests that QNS of $p_T$-spectra found in~\cite{Song:2017gcz, Zhang:2018vyr} is not only valid in $pA$ and $pp$ but also in $AA$ collisions and hence is a universal property of $p_T$-spectra of hadrons in all three different kinds of hadronic reactions.
 The QNS is a direct consequence of quark combination under the rule of equal velocity combination (EVC).
It provides a convenient way of extracting the $p_{T}$-distribution of quarks from the data for those of hadrons.
We extracted in this paper the $p_T$-spectrum of strange quarks from the data of $\Omega$
and $\phi$ and studied its properties such as the average $p_T$ and the radial flow velocity of strange quarks within the blast-wave model.
We show that this may get deep insights into the properties of QGP
and/or mechanisms of hadronic interactions in high energy collisions.

As discussions, we would like also to mention that QNS might be used as a test of different mechanisms.
As mentioned, quark combinations under EVC provide the most direct and natural explanation.
In contrast, we have also checked that, with default parameters,
event generators where the fragmentation mechanism is adopted
such as PYTHIA~\cite{Sjostrand:2014zea}, Herwig~\cite{Bahr:2008pv}, AMPT~\cite{Lin:2004en}
and HIJING~\cite{Wang:1991hta} show significant deviations from QNS.
Including color re-connection and/or string overlap effects~\cite{Ortiz:2013yxa,Bierlich:2014xba}
does not significantly improve the case.
The QNS might also provide important constraints on details of quark combination mechanism.
EVC at the constituent quark level provides the most natural explanations
while others such as the Wigner wave function method in coalescence models~\cite{Greco:2003xt,Fries:2003kq,Chen:2003tn,Chen:2006vc,Pu:2018eei},
the parton recombination~\cite{Hwa:2004ir} with recombination functions determined by the valon model~\cite{Hwa:1979pn,Hwa:2002mv},
and AMPT with string melting that adopts a coalescence mechanism via finite combination radius~\cite{Ye:2017ooc} and so on
seem all slightly deviate from such precise QNS for $p_T$-spectra of the produced hadrons.
Further studies along this direction, both experimentally and theoretically, should be worthwhile and encouraging.

We emphasize once more that the purpose of this paper is to study 
whether QNS discussed in \cite{Song:2017gcz, Zhang:2018vyr} for $pp$ and $pA$ is also valid in $AA$ collisions. 
For this purpose, $\Omega$ and $\phi$ production are the best examples 
and it is very fortunate that we have also data available for the productions of these two hadrons. 
We have shown that QNS is indeed also valid in this case and 
we have presented a possible explanation of QNS and examples of further applications in connection to properties of QGP. 
In \cite{Song:2017gcz, Zhang:2018vyr}, we have checked that such QNS is valid in $pp$ and $pA$ not only for hadrons composed 
of strange quarks but also for those composed of $u$ and/or $d$ quarks. 
It is also very interesting to check whether this is also the case for $AA$ collisions when corresponding data are available. 
\begin{acknowledgments}
We thank S. Y. Li and Q. Wang for helpful discussions. 
We thank X.L. Zhu for providing us the latest data of $\Omega^{-}$ in Au+Au collisions at $\sqrt{s_{NN}}=$ 200 GeV. 
This work was supported in part by the National Natural Science Foundation of China 
under Grant Nos. 11975011, 11675092 and 11890713,  Project of Shandong Province
Higher Educational Science and Technology Program (J18KA228), and
Shandong Province Natural Science Foundation Grant No. ZR2019YQ06 and ZR2019MA053. 
\end{acknowledgments}



\bibliographystyle{apsrev4-1}
\bibliography{ref}

\end{document}